\documentclass[conference]{IEEEtran}
\IEEEoverridecommandlockouts

\usepackage{cite}
\usepackage{amsmath,amssymb,amsfonts}
\usepackage{graphicx}
\usepackage{textcomp}
\usepackage{xcolor}
\usepackage{hyperref} 
\usepackage{url} 
\usepackage{booktabs} 
\usepackage{amsfonts} 
\usepackage{nicefrac} 
\usepackage{longtable} 
\usepackage{array} 
\usepackage{multirow} 
\usepackage{soul} 
\usepackage{amsmath} 
\usepackage{enumitem} 
\usepackage{algorithm}
\usepackage{algpseudocode}

\newcommand{\MyModel}{M-RAG}

\def\BibTeX{{\rm B\kern-.05em{\sc i\kern-.025em b}\kern-.08em
    T\kern-.1667em\lower.7ex\hbox{E}\kern-.125emX}}
\begin{document}

\title{M-RAG: Semantic Key-Value Indexing for Retrieval-Augmented Generation}

\author{\IEEEauthorblockN{Xu Sun\textsuperscript{1}, Tongkai Xu\textsuperscript{1}, Baiheng Xie\textsuperscript{1}, Jun Zhang\textsuperscript{1}, 
Li Huang\textsuperscript{1,*}, Qiang Gao\textsuperscript{1}, Kunpeng Zhang\textsuperscript{2}}
\IEEEauthorblockA{\textsuperscript{1}School of Computing and Artificial Intelligence, 
Southwestern University of Finance and Economics, China\\
\textsuperscript{2}Information Systems, 
University of Maryland, United States\\
\{42123079,42311038,42211042,225085400006\}@smail.swufe.edu.cn, \\
\{lihuang,qianggao\}@swufe.edu.cn,
kpzhang@umd.edu}
\thanks{\textsuperscript{*}Corresponding author: Li Huang (lihuang@swufe.edu.cn).}
}

\maketitle

\begin{abstract}
Retrieval-augmented generation (RAG) turns external documents into evidence for large language models. In practice, this is also a data access problem: a system must decide what to index, what to retrieve, and what evidence to place in the context under a token budget. Most RAG pipelines use text chunks for both lookup and generation. This couples two different objectives. Retrieval benefits from compact and discriminative records, while generation needs contextual and faithful evidence. As a result, small chunks may fragment answer-bearing information, whereas large chunks may introduce noise and waste the context budget. We propose \MyModel, a semantic key-value indexing layer for budget-constrained RAG query processing. \MyModel~extracts meta-markers from complete documents, where each record contains a retrieval key, an information value, and provenance pointers. Online retrieval operates over the key field, which can be searched by dense vector retrieval or sparse lexical retrieval; the paired values are returned as generation payloads and assembled under the token budget. Provenance pointers further support coverage validation and position-aware context ordering. This design separates the physical index entry from the evidence payload without changing the underlying retriever or generator. Experiments on LongBench QA subtasks show that \MyModel~achieves competitive or better accuracy than representative chunk-based baselines, especially under tight token budgets. Further analyses show high document coverage, stronger robustness under expanding candidate corpora, and lower online retrieval latency. These results suggest that semantic key-value indexing is a practical access method for RAG workloads.
\end{abstract}


\section{Introduction}
\label{sec:introduction}
Retrieval-augmented generation (RAG) has become a common interface between large language models (LLMs) and external data. In data-intensive settings, such as enterprise knowledge bases, scientific literature search, legal analysis~\cite{louis2024interpretable}, and financial reporting~\cite{zhang2023enhancing}, a RAG system depends on the relevance, completeness, and organization of the evidence it presents to the generator. This makes RAG not only a generation problem, but also a data management problem: how should evidence be represented, indexed, retrieved, and assembled before an LLM reads it?

From this perspective, RAG is an indexing and query processing problem under a token budget. Given a document collection and a user query, the system must support selective lookup, low-latency retrieval, compact context assembly, and traceability to the source data. The retrieved records are not the final output. They are intermediate evidence consumed by an LLM, whose context window is limited and whose generation can be affected by irrelevant or poorly organized context. A RAG access method therefore has to balance retrieval accuracy, evidence fidelity, provenance, and online efficiency.

\begin{figure}[t]
\centering
\includegraphics[width=0.85\linewidth]{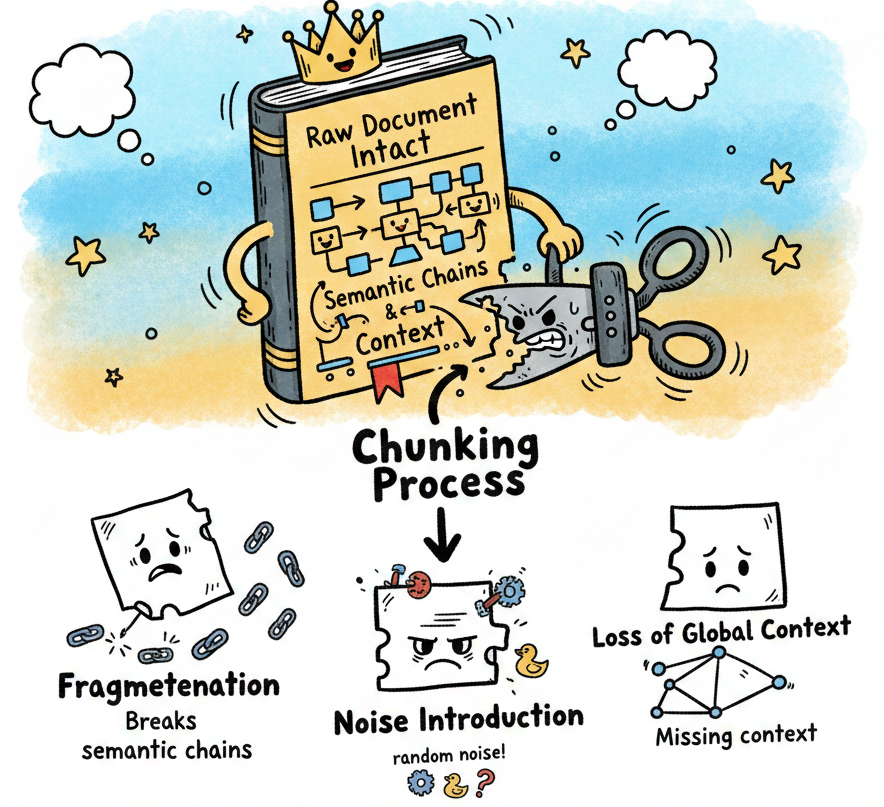}
\caption{An illustration of information fragmentation and structural loss caused by chunk-based RAG.}
\vspace{-4mm}
\label{fig:chunking}
\end{figure}

Most RAG systems instantiate this access method with text chunks. A document is split into fixed-size windows, overlapping windows, semantic segments, or structure-aware spans. At query time, the retriever selects several chunks, and the selected chunks are inserted into the LLM context~\cite{lewis2020retrieval,guu2020realm,kamradt2024semantic,gao2024retrievalaugmentedgenerationlargelanguage}. This design is simple, scalable, and compatible with standard vector indexes. It also makes a strong physical-design choice: the object used as the index entry is the same object returned as the generation payload.

This choice is convenient, but the two roles have different objectives. Retrieval benefits from compact and discriminative records that match a query precisely and avoid mixing unrelated topics. Generation benefits from complete and contextual evidence that preserves entities, relations, dates, discourse structure, and other answer-bearing details. A single chunk is asked to serve both roles. Small chunks may improve matching but fragment evidence. Large chunks preserve more context but introduce noise, waste token budget, and enlarge the online search space. Overlap reduces some boundary errors, but it increases redundancy and keeps the same coupling between \emph{index entry} and \emph{payload}. Figure~\ref{fig:chunking} illustrates this failure mode: chunk boundaries can split answer-bearing information and weaken the document structure seen by the retriever and generator. The issue is therefore not only where to cut a document; it is that chunking exposes a fixed access interface in which lookup representation and generation payload are tied together.

Recent work has made chunking more adaptive. Semantic chunking groups adjacent text by similarity. PIC~\cite{wang-etal-2025-document} uses document summaries as pseudo-instructions for adaptive grouping. MoC~\cite{zhao-etal-2025-moc} learns mixtures of chunking strategies. DOS RAG~\cite{laitenberger-etal-2025-stronger} shows that preserving the original document structure can be a strong baseline. These methods improve segmentation quality and demonstrate that the retrieval unit matters. However, they still retain the same access interface: retrieve a text span, then use that span as generation context. Long-context LLMs do not remove this design question. Although longer context windows reduce the need to use retrieval only as external memory, they still require systems to decide which evidence should enter the context, and they remain sensitive to irrelevant or distracting information~\cite{hsieh2024ruler,wang-etal-2024-retrieve,asai2024selfrag,nuengsigkapian2025hifiraghierarchicalcontentfiltering,wang2025retrievalaugmentedquestionanswering}. Recent chunk-free retrieval work, such as CFIC~\cite{qian-etal-2024-grounding}, further suggests that raw chunks need not be the basic retrieval object.

This paper asks a simple question: should the \emph{retrieval representation} be the same text as the \emph{generation payload}? We argue that it should not. We propose \MyModel, a semantic key-value indexing layer for RAG. Instead of indexing raw chunks, \MyModel~extracts semantic records, called meta-markers, from complete documents. Each meta-marker contains three fields: a retrieval key $k$, an information value $v$, and provenance pointers $p$. The key is a compact, query-facing description used for lookup. The value is a richer evidence payload used for generation. The provenance pointers connect the record back to source paragraphs. For example, a key may describe an information need such as a definition, a date, a relation, or an event, while the paired value stores the answer-bearing evidence and the provenance pointers identify where that evidence comes from.

\MyModel~indexes keys and returns values. At query time, retrieval operates over the key field rather than the value field. The key can be embedded for dense retrieval or indexed as text for sparse retrieval, such as BM25. Once a key is retrieved, its paired value is assembled into the generation context under a token budget. This separates the lookup representation from the returned payload. The key can be optimized for precise matching, while the value can be optimized for evidence fidelity. The provenance pointers support coverage validation and position-aware context ordering. In this sense, \MyModel~applies a familiar database principle to RAG: keep the index lightweight, and keep the payload faithful to the data needed by downstream operators.

\MyModel~is designed as a modular indexing layer. Offline, an off-the-shelf LLM extracts meta-markers from tagged documents. A coverage validation step checks whether the extracted records cover the source document and creates conservative fallback records for uncovered units. Online, the system retrieves relevant keys through a dense or sparse index, selects paired values under a token budget, and orders the selected values by source position or similarity before generation. Retrieval operates over compact semantic keys; generation reads contextual evidence values.

Our contributions are:
\begin{itemize}
    \item We formulate RAG retrieval as a semantic indexing and query processing problem, and identify the coupling between index entry and generation payload as a source of the chunk-size tradeoff.
    \item We introduce a semantic key-value marker representation for RAG. It separates retrieval keys, information values, and provenance pointers, allowing lookup, generation, and traceability to be handled by distinct fields.
    \item We design an offline construction and validation pipeline together with an online budget-aware query processing procedure for key-based retrieval and value-based context assembly.
    \item We evaluate \MyModel~on LongBench QA subtasks. The results show competitive or higher accuracy under tight token budgets, high document coverage, and lower online retrieval latency than representative chunk-based baselines.
\end{itemize}

\section{Related Work}
Retrieval-augmented generation (RAG) connects language models with external corpora by retrieving evidence before generation~\cite{lewis2020retrieval,guu2020realm}. The basic interface has since evolved into dense retrieval pipelines, modular RAG systems, graph-based retrieval, and agentic workflows~\cite{fan2024survey,gao2024retrievalaugmentedgenerationlargelanguage,singh2025agenticretrievalaugmentedgenerationsurvey,li-etal-2025-survey}. Despite these architectural differences, RAG systems share a core data-management question: what representation should be indexed, and what evidence should be returned to the generator?

\noindent \textbf{Indexing and access methods for semantic data.}
Data systems typically separate logical records from physical access paths. In semantic search, dense-vector indexes such as HNSW~\cite{malkov2018efficient} and sparse lexical indexes provide alternative lookup paths over a collection. This view is useful for RAG because chunking is not only preprocessing; it determines the physical record exposed to the access path. \MyModel~follows this access-method perspective by materializing a record layout in which the field used for lookup is different from the payload returned to the downstream operator.

\noindent \textbf{Retrieval units in RAG.}
Most RAG systems instantiate this choice with text spans. Fixed-size, overlapping, semantic, and structure-aware chunking methods differ in how they determine document boundaries, but they largely preserve the same access interface: a query is matched against a text unit, and the selected unit is inserted into the generation context. Prior work has shown that segmentation quality affects retrieval and downstream QA, and that the choice of granularity creates a tension between noise and coherence~\cite{wang2025retrievalsucceedsfailsrethinking,zhao-etal-2025-moc}. PIC uses document summaries as pseudo-instructions for adaptive grouping~\cite{wang-etal-2025-document}, while MoC learns mixtures of chunking learners and applies the learned chunkers to extract text spans~\cite{zhao-etal-2025-moc}. Proposition-level dense retrieval studies a finer granularity by indexing concise factual propositions rather than passages or sentences~\cite{chen-etal-2024-dense}. These methods show that retrieval-unit design is important, but they mainly optimize the boundary or granularity of the text object being retrieved.

\noindent \textbf{Structure-aware and graph-enhanced RAG.}
Beyond segmentation, recent work has emphasized the organization of evidence. DOS RAG preserves the original document structure and passage order as a simple retrieve-then-read baseline for long-context language models~\cite{laitenberger-etal-2025-stronger}. Graph-augmented methods construct entity-relation or community structures over indexed text to support retrieval beyond flat chunk collections~\cite{edge2025localglobalgraphrag,guo-etal-2025-lightrag}. LightRAG, for example, incorporates graph structures into text indexing and retrieval and combines them with vector representations~\cite{guo-etal-2025-lightrag}. AGRAG further improves graph-based RAG by using statistics-based entity extraction and retrieving minimum-cost maximum-influence reasoning subgraphs~\cite{wang2026agragadvancedgraphbasedretrievalaugmented}. These methods model structural dependencies in the evidence space. They are complementary to \MyModel: rather than proposing a new graph retriever or generator, \MyModel~changes the logical record layout exposed to retrieval. The resulting records can still be searched with dense, sparse, or graph-enhanced backends.

\noindent \textbf{Decoupled and chunk-free representations.}
The closest line of work relaxes the assumption that raw chunks must be the basic retrieval object. CFIC proposes chunking-free in-context retrieval by using encoded document hidden states and decoding strategies to identify evidence text for a query~\cite{qian-etal-2024-grounding}. HeteRAG studies decoupled knowledge representations, using short chunks for generation and context-enriched multi-granular views for retrieval~\cite{yang2025heterag}. These studies are closely related because they also separate, to different degrees, the representation used for retrieval from the evidence consumed by generation. \MyModel~takes a database access-method perspective on this design space. It materializes a persistent semantic key-value record layout: a compact query-facing key is indexed for lookup, a richer value is returned for generation, and provenance pointers keep the record traceable to source units. This layout explicitly separates query matching from evidence delivery while preserving the standard RAG generator interface.

\section{Methodology}
\label{sec:MRAG}
\begin{figure*}[htbp]
\centering
\includegraphics[width=\textwidth]{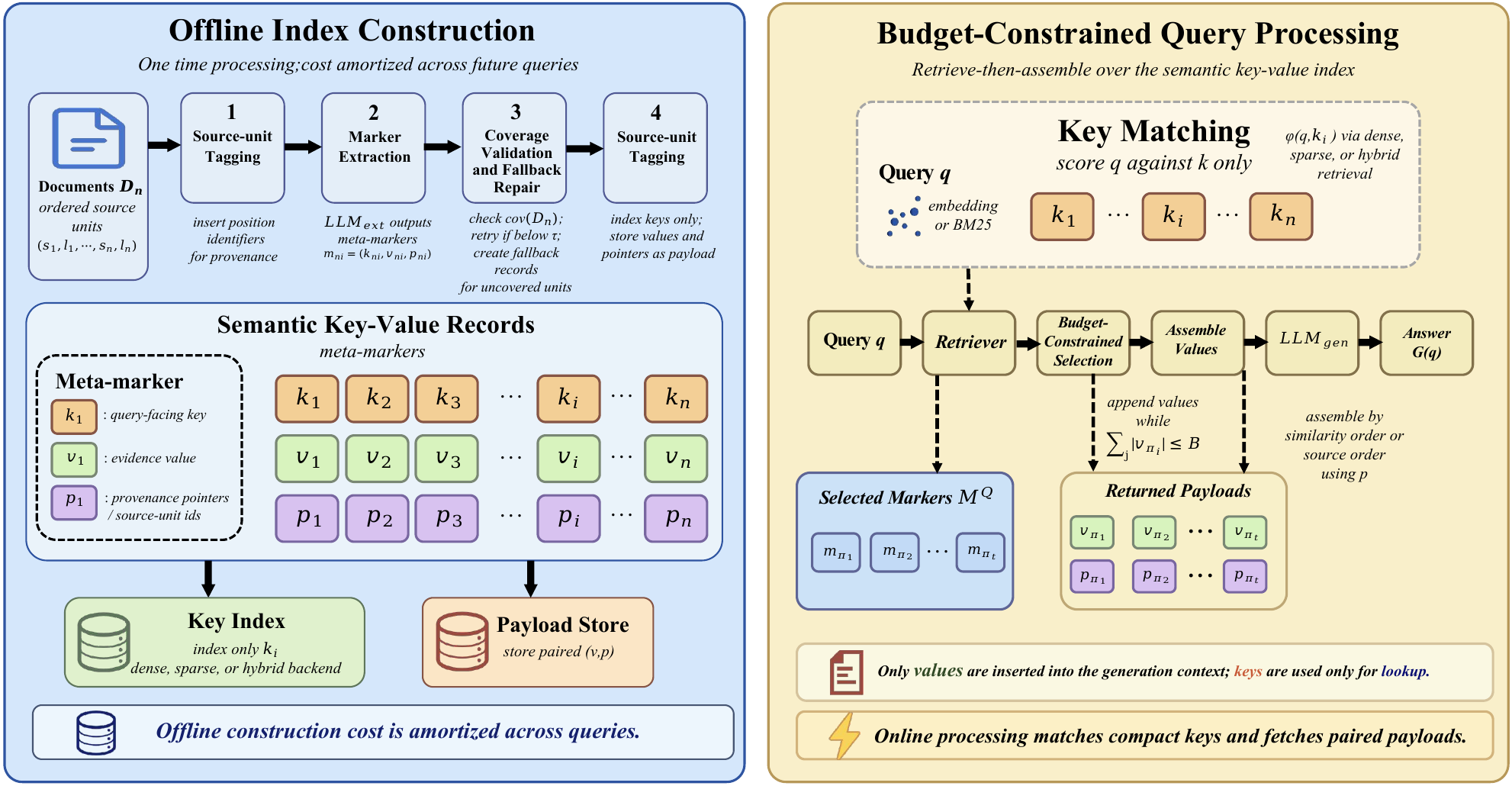}
\vspace{-5mm}
\caption{The overall architecture of~\MyModel.}
\label{fig:m-rag-framework}
\vspace{-4mm}
\end{figure*}

\MyModel~treats RAG retrieval as an access-method design problem. A RAG system needs records that are selective enough for lookup, but also informative enough for generation. Chunk-based RAG uses the same physical text span for both roles. \MyModel~separates these roles by materializing documents into semantic key-value records. As shown in Figure~\ref{fig:m-rag-framework}, the system constructs the records offline and, at query time, retrieves by semantic keys while returning paired evidence values to the generator.

\subsection{Semantic Key-Value Record Model}

We consider a document collection $\mathcal{D}=\{D_1,\ldots,D_N\}$. Each document $D_n$ is represented as an ordered sequence of source units,
\begin{equation}
    D_n=(s_{n,1},\ldots,s_{n,L_n}),
\end{equation}
where a source unit can be a paragraph or a token-based segment. Source units are not the logical retrieval records in \MyModel. They serve as the physical units used for provenance, coverage validation, and source-order reconstruction.

\MyModel~materializes each document into a set of semantic key-value records, called meta-markers:
\begin{align}
    \mathbf{M}_{D_n} &= \{m_{n,1},\ldots,m_{n,|\mathbf{M}_{D_n}|}\},\\
    m_{n,i} &= (k_{n,i}, v_{n,i}, p_{n,i}).
\end{align}
Here $k_{n,i}$ is a query-facing key, $v_{n,i}$ is the evidence value returned for generation, and $p_{n,i}\subseteq\{1,\ldots,L_n\}$ is a set of provenance pointers to source units in $D_n$. The key is the indexed object, the value is the returned object, and the pointer keeps the record traceable to the source document. The global marker collection is
\begin{equation}
    \mathcal{M}=\bigcup_{n=1}^{N}\mathbf{M}_{D_n}.
\end{equation}

Given a query $q$ and a token budget $B$, the online task is to select a sequence of markers from $\mathcal{M}$ and assemble their values as the external context:
\begin{align}
    \mathbf{M}^{q} &= (m_{\pi_1},\ldots,m_{\pi_t}),\\
    \mathrm{s.t.}\quad & \sum_{j=1}^{t} |v_{\pi_j}| \leq B.
\end{align}
Only the keys $k_{\pi_j}$ are used to estimate relevance. Only the values $v_{\pi_j}$ are placed in the generation context:
\begin{equation}
    G(q)=\operatorname{LLM}_{\mathrm{gen}}(q, v_{\pi_1}\oplus\cdots\oplus v_{\pi_t}),
\end{equation}
where $\oplus$ denotes textual concatenation. This record model separates the lookup representation from the generation payload while preserving a link to the original source units.

\subsection{Offline Index Construction}

\MyModel~builds the semantic key-value index before serving queries. The construction pipeline has four steps: source-unit tagging, marker extraction, coverage validation with fallback repair, and key indexing.

\noindent \textbf{Source-unit tagging.}
For each document $D_n$, every source unit $s_{n,j}$ is assigned a position identifier. These identifiers are inserted into the extractor input so that generated records can refer back to the units from which their evidence is derived. The identifiers are not used as retrieval content; they provide provenance for validation and later context ordering.

\noindent \textbf{Marker extraction.}
Given a tagged document and an extraction instruction $\mathbb{P}$, an off-the-shelf LLM produces a set of records:
\begin{equation}
    \mathbf{M}_{D_n}=\operatorname{LLM}_{\mathrm{ext}}(D_n,\mathbb{P}).
\end{equation}
The extractor is constrained to produce records with a query-facing key, a self-contained evidence value, and bounded provenance pointers. The prompt is used only to specify this output schema and the extraction constraints; the complete prompt and implementation details are provided in the accompanying artifact. Table~\ref{tab:prompt-structure} summarizes the prompt structure used by \MyModel.

\begin{table}[t]
\centering
\caption{Prompt structure for marker extraction.}
\label{tab:prompt-structure}
\resizebox{\linewidth}{!}{
\begin{tabular}{l p{5.8cm}}
\toprule
Component & Role \\
\midrule
Input & Tagged source units with position identifiers and an expected marker count derived from document length. \\
Output schema & A JSON-style list of records, each with key $k$, value $v$, and provenance pointer $p$. \\
Granularity & Prefer fine-grained records; each record is constrained to a small bounded set of source units. \\
Value field & Produce a self-contained evidence payload that preserves entities, attributes, relations, dates, and other answer-bearing details. \\
Key field & Produce a query-facing description that summarizes and retrieves the paired value. \\
Pointer field & Return source-unit identifiers used to construct the value. \\
Coverage & Cover all source units, allowing overlap across records; uncovered units are handled by fallback repair. \\
\bottomrule
\end{tabular}
}
\vspace{-6mm}
\end{table}

\noindent \textbf{Coverage validation and fallback repair.}
Because the records are generated by an LLM, \MyModel~validates the extracted index before exposing it to online retrieval. For a document $D_n$, provenance coverage is defined as
\begin{equation}
    \operatorname{cov}(D_n)=
    \frac{\left|\bigcup_{m\in \mathbf{M}_{D_n}} p(m)\right|}{L_n}.
\end{equation}
If $\operatorname{cov}(D_n)$ is below a threshold $\tau$, extraction is retried. If the retries still leave uncovered source units, \MyModel~applies a conservative repair: each uncovered source unit is converted into a fallback record whose key and value are both the original unit content, and whose pointer refers to that unit. This repair prevents LLM extraction failures from making any source unit unreachable.

\noindent \textbf{Key indexing.}
After validation, only keys are indexed. The values and provenance pointers are stored as payload fields attached to the corresponding records. In a dense implementation, the key set
\begin{equation}
    K=\{k_i \mid m_i=(k_i,v_i,p_i)\in\mathcal{M}\}
\end{equation}
is embedded by a sentence embedding model $\mathbb{E}(\cdot)$ and stored in a vector access path:
\begin{equation}
    \mathbf{e}_{k_i}=\mathbb{E}(k_i).
\end{equation}
In a sparse implementation, the same key strings can be indexed by a lexical retriever such as BM25. In both cases, online matching operates over keys rather than over the richer evidence values.

\subsection{Budget-Constrained Query Processing}

At query time, \MyModel~executes a retrieve-then-assemble plan over the key-value index. The retrieval backend ranks records by applying a scoring function to the query and the key field:
\begin{equation}
    \operatorname{score}(q,m_i)=\phi(q,k_i),
\end{equation}
where $\phi$ can be implemented as dense embedding similarity, lexical BM25 scoring, or another retrieval function over keys. For dense retrieval, the query is embedded as $\mathbf{e}_q=\mathbb{E}(q)$ and submitted to the vector access path, which returns approximate nearest-neighbor keys. For sparse retrieval, $q$ is matched against the lexical index over key strings.

Let $(m_{\pi_1},m_{\pi_2},\ldots)$ be the candidate sequence returned by the backend and sorted by decreasing key score. \MyModel~scans this sequence and appends the paired value $v_{\pi_j}$ while the cumulative value length remains within the token budget:
\begin{equation}
    \sum_{j=1}^{t}|v_{\pi_j}| \leq B
    \quad\text{and}\quad
    \sum_{j=1}^{t+1}|v_{\pi_j}| > B.
\end{equation}
The selected values can then be assembled by similarity order, which preserves the retrieval ranking, or by source order, which uses the provenance pointers to recover the document order. The assembled values form the context consumed by the generator. The retrieved keys are not inserted into the generation context unless they are also part of the stored value.

\subsection{Design Analysis}

The design changes the physical access interface of RAG. In chunk-based systems, a text span must be selective enough to match a query and complete enough to support generation. These requirements are often in tension: shorter spans improve specificity but can fragment evidence, while longer spans preserve context but mix topics and spend more of the token budget. \MyModel~assigns these requirements to different fields. The key provides a compact lookup representation, while the value carries the evidence consumed downstream.

The provenance field does not participate in ranking, but it makes the semantic index auditable. Before indexing, provenance pointers support a coverage check over the source units and allow fallback repair for uncovered units. After retrieval, the same pointers support source-order assembly, which is useful when answer evidence depends on document structure rather than only on local similarity.

The main cost of this design is moved to the offline construction phase. Marker extraction materializes semantic records before queries arrive, while online processing only matches keys and fetches paired payloads. This division is suitable for read-heavy RAG workloads, where a constructed index is reused across many queries. It also keeps the retrieval backend modular: dense, sparse, or hybrid retrievers can operate over the same key field without changing the generator interface.

\section{Experiments}
\label{sec:experiment}
\subsection{Experimental Setup}
\label{subsec:setup}

\noindent \textbf{Benchmark.} We evaluate \MyModel~on QA subtasks from LongBench~\cite{bai2024longbenchbilingualmultitaskbenchmark}, a benchmark for long-context understanding. The evaluation uses NarrativeQA~\cite{kocisky-etal-2018-narrativeqa} and Qasper~\cite{dasigi-etal-2021-dataset} for single-hop QA, and 2WikiMultihopQA~\cite{ho-etal-2020-constructing} for multi-hop QA. Each subtask contains 200 samples and covers contexts longer than 3K tokens on average, spanning narrative, scientific, and multi-document reasoning workloads.

\noindent \textbf{Baselines.} We compare \MyModel~with representative retrieval-unit construction methods. The chunk-based baselines include Fixed-Size RAG with 128-token segments, Fixed-Size RAG with overlap, Semantic RAG~\cite{kamradt2024semantic}, PIC RAG~\cite{wang-etal-2025-document}, and DOS RAG~\cite{laitenberger-etal-2025-stronger}. These methods address the same physical-design question as \MyModel: how evidence should be organized before retrieval and generation. We also include LightRAG~\cite{guo-etal-2025-lightrag} as a graph-enhanced RAG baseline that combines graph structures with text retrieval.

\noindent \textbf{Retrieval and generation.} We evaluate each method under two retrieval backends. For dense retrieval, we use BAAI/bge-m3~\cite{bge-m3}. For sparse retrieval, we use BM25 over each method's retrieval field: semantic keys for \MyModel~and text chunks for chunk-based baselines. All methods are evaluated under the same retrieved-evidence budgets of 128$\times$1, 128$\times$3, and 128$\times$5 tokens. Generation is performed by Qwen3-30B-A3B-Instruct-2507~\cite{qwen3technicalreport} with the same answer-generation prompt across methods. The generator receives the selected information values for \MyModel~and the retrieved text chunks for chunk-based baselines.

\noindent \textbf{\MyModel~configuration.} Marker extraction uses DeepSeek-V3.2~\cite{deepseekai2025deepseekv32pushingfrontieropen} with one prompt template across all benchmarks. We report zero-shot and few-shot extraction variants; in the few-shot setting, the extractor receives one benchmark-specific marker example following the same key-value-pointer schema. We also evaluate two context assembly orders: position sorting (-P), which orders values by provenance position, and similarity sorting (-S), which preserves retrieval ranking. Both the extractor and generator use temperature 0. Each experiment is repeated five times, and we report mean and standard deviation. Experiments are conducted on a server with NVIDIA A800 GPUs (80GB). The implementation, prompts, few-shot examples, and reproduction artifacts are available at \url{https://github.com/9sxx/M-RAG}.

\subsection{End-to-End QA Performance}
\label{subsec:end-to-end-performance}

\begin{table*}[t]
\centering
\caption{Performance comparison in F1 score following LongBench. Results are grouped by dataset. For each method, we report performance under both \textbf{BAAI/bge-m3} and \textbf{BM25} retrievers across three retrieval budgets (128$\times$1 / 128$\times$3 / 128$\times$5). Best result in \textbf{bold}, second best \underline{underlined}. \MyModel~reports zero-shot and few-shot variants with position sorting (-P) and similarity sorting (-S).}
\label{tab:m-rag-results}
\begin{tabular}{l ccc ccc}
\toprule
 & \multicolumn{3}{c}{\textbf{BAAI/bge-m3}} & \multicolumn{3}{c}{\textbf{BM25}} \\
\cmidrule(lr){2-4} \cmidrule(lr){5-7}
 & 128$\times$1 & 128$\times$3 & 128$\times$5 & 128$\times$1 & 128$\times$3 & 128$\times$5 \\
\midrule
\multicolumn{7}{c}{\textbf{NarrativeQA}} \\
\midrule
Fixed-Size & 0.0660 $\pm$ 0.0000 & 0.0952 $\pm$ 0.0002 & 0.1094 $\pm$ 0.0001 & 0.0568 $\pm$ 0.0015 & 0.0920 $\pm$ 0.0032 & 0.1178 $\pm$ 0.0009 \\
\quad w. overlap & 0.0581 $\pm$ 0.0000 & 0.0906 $\pm$ 0.0000 & 0.0982 $\pm$ 0.0001 & 0.0545 $\pm$ 0.0015 & 0.0877 $\pm$ 0.0050 & \textbf{0.1201 $\pm$ 0.0040} \\
Semantic & 0.0617 $\pm$ 0.0000 & 0.1200 $\pm$ 0.0002 & 0.1449 $\pm$ 0.0009 & 0.0574 $\pm$ 0.0015 & 0.0940 $\pm$ 0.0029 & 0.1102 $\pm$ 0.0016 \\
PIC & 0.0618 $\pm$ 0.0009 & 0.1092 $\pm$ 0.0001 & 0.1226 $\pm$ 0.0005 & 0.0565 $\pm$ 0.0030 & 0.0962 $\pm$ 0.0029 & 0.1148 $\pm$ 0.0086 \\
DOS & \textbf{0.0843 $\pm$ 0.0000} & 0.1191 $\pm$ 0.0000 & 0.1398 $\pm$ 0.0001 & 0.0434 $\pm$ 0.0038 & 0.0926 $\pm$ 0.0042 & 0.1146 $\pm$ 0.0033 \\
LightRAG & 0.0614 $\pm$ 0.0015 & 0.1051 $\pm$ 0.0076 & 0.1201 $\pm$ 0.0042 & 0.0027 $\pm$ 0.0010 & 0.0412 $\pm$ 0.0037 & 0.0700 $\pm$ 0.0046 \\
\midrule
\textbf{\MyModel-P} & 0.0684 $\pm$ 0.0006 & \textbf{0.1279 $\pm$ 0.0018} & \underline{0.1459 $\pm$ 0.0022} & 0.0572 $\pm$ 0.0041 & 0.0940 $\pm$ 0.0007 & 0.1176 $\pm$ 0.0045 \\
\textbf{\MyModel-S} & 0.0657 $\pm$ 0.0025 & \underline{0.1214 $\pm$ 0.0014} & \textbf{0.1476 $\pm$ 0.0031} & 0.0585 $\pm$ 0.0013 & 0.0984 $\pm$ 0.0019 & 0.1183 $\pm$ 0.0040 \\
\textbf{\MyModel$^\dag$-P} & \underline{0.0736 $\pm$ 0.0029} & 0.1247 $\pm$ 0.0030 & 0.1381 $\pm$ 0.0018 & \textbf{0.0771 $\pm$ 0.0039} & \textbf{0.1078 $\pm$ 0.0053} & 0.1147 $\pm$ 0.0024 \\
\textbf{\MyModel$^\dag$-S} & 0.0644 $\pm$ 0.0004 & 0.1208 $\pm$ 0.0029 & 0.1357 $\pm$ 0.0013 & \underline{0.0757 $\pm$ 0.0014} & \underline{0.1057 $\pm$ 0.0025} & \underline{0.1197 $\pm$ 0.0015} \\
\midrule
\multicolumn{7}{c}{\textbf{Qasper}} \\
\midrule
Fixed-Size & 0.1316 $\pm$ 0.0000 & 0.1908 $\pm$ 0.0000 & 0.2425 $\pm$ 0.0006 & 0.1043 $\pm$ 0.0037 & 0.1693 $\pm$ 0.0081 & 0.2221 $\pm$ 0.0015 \\
\quad w. overlap & 0.1294 $\pm$ 0.0003 & 0.1963 $\pm$ 0.0002 & 0.2361 $\pm$ 0.0014 & 0.1026 $\pm$ 0.0076 & 0.2032 $\pm$ 0.0080 & 0.2257 $\pm$ 0.0067 \\
Semantic & 0.1654 $\pm$ 0.0008 & \underline{0.2482 $\pm$ 0.0007} & 0.2683 $\pm$ 0.0003 & 0.1297 $\pm$ 0.0102 & 0.1769 $\pm$ 0.0034 & 0.2267 $\pm$ 0.0032 \\
PIC & 0.1581 $\pm$ 0.0006 & \textbf{0.2530 $\pm$ 0.0036} & \textbf{0.2838 $\pm$ 0.0021} & 0.1230 $\pm$ 0.0075 & 0.1939 $\pm$ 0.0046 & 0.2341 $\pm$ 0.0066 \\
DOS & 0.1643 $\pm$ 0.0009 & 0.2269 $\pm$ 0.0009 & 0.2635 $\pm$ 0.0024 & 0.1142 $\pm$ 0.0038 & 0.1798 $\pm$ 0.0036 & \textbf{0.2462 $\pm$ 0.0064} \\
LightRAG & 0.0858 $\pm$ 0.0010 & 0.1761 $\pm$ 0.0050 & 0.1928 $\pm$ 0.0051 & 0.0362 $\pm$ 0.0016 & 0.0761 $\pm$ 0.0038 & 0.1344 $\pm$ 0.0062 \\
\midrule
\textbf{\MyModel-P} & 0.1436 $\pm$ 0.0018 & 0.2317 $\pm$ 0.0013 & 0.2456 $\pm$ 0.0013 & \textbf{0.1364 $\pm$ 0.0059} & \textbf{0.2047 $\pm$ 0.0031} & 0.2268 $\pm$ 0.0054 \\
\textbf{\MyModel-S} & 0.1390 $\pm$ 0.0037 & 0.2308 $\pm$ 0.0032 & 0.2655 $\pm$ 0.0055 & 0.1159 $\pm$ 0.0030 & 0.2008 $\pm$ 0.0164 & 0.2404 $\pm$ 0.0053 \\
\textbf{\MyModel$^\dag$-P} & \underline{0.1806 $\pm$ 0.0043} & 0.2370 $\pm$ 0.0037 & \underline{0.2693 $\pm$ 0.0061} & 0.1174 $\pm$ 0.0018 & \underline{0.2038 $\pm$ 0.0054} & 0.2389 $\pm$ 0.0029 \\
\textbf{\MyModel$^\dag$-S} & \textbf{0.1820 $\pm$ 0.0025} & 0.2224 $\pm$ 0.0041 & 0.2690 $\pm$ 0.0015 & \underline{0.1298 $\pm$ 0.0025} & 0.2026 $\pm$ 0.0068 & \underline{0.2412 $\pm$ 0.0020} \\
\midrule
\multicolumn{7}{c}{\textbf{2WikiMultihopQA}} \\
\midrule
Fixed-Size & 0.2220 $\pm$ 0.0000 & 0.3239 $\pm$ 0.0038 & 0.3832 $\pm$ 0.0023 & \textbf{0.2008 $\pm$ 0.0153} & 0.2906 $\pm$ 0.0096 & 0.3231 $\pm$ 0.0161 \\
\quad w. overlap & 0.1792 $\pm$ 0.0000 & 0.3186 $\pm$ 0.0004 & 0.3740 $\pm$ 0.0018 & 0.1793 $\pm$ 0.0115 & 0.2898 $\pm$ 0.0062 & 0.3058 $\pm$ 0.0040 \\
Semantic & 0.1875 $\pm$ 0.0005 & 0.3125 $\pm$ 0.0030 & 0.3831 $\pm$ 0.0013 & 0.1693 $\pm$ 0.0125 & 0.2881 $\pm$ 0.0184 & 0.3312 $\pm$ 0.0057 \\
PIC & 0.2066 $\pm$ 0.0000 & 0.3236 $\pm$ 0.0015 & 0.3947 $\pm$ 0.0026 & 0.1610 $\pm$ 0.0084 & 0.2886 $\pm$ 0.0111 & 0.2993 $\pm$ 0.0100 \\
DOS & 0.2023 $\pm$ 0.0000 & 0.3255 $\pm$ 0.0004 & \textbf{0.4110 $\pm$ 0.0028} & \underline{0.1908 $\pm$ 0.0021} & \underline{0.2991 $\pm$ 0.0200} & 0.3245 $\pm$ 0.0123 \\
LightRAG & 0.1308 $\pm$ 0.0077 & 0.2304 $\pm$ 0.0040 & 0.2687 $\pm$ 0.0057 & 0.0604 $\pm$ 0.0062 & 0.1063 $\pm$ 0.0042 & 0.1756 $\pm$ 0.0047 \\
\midrule
\textbf{\MyModel-P} & 0.2134 $\pm$ 0.0016 & 0.3229 $\pm$ 0.0030 & 0.3494 $\pm$ 0.0057 & 0.1657 $\pm$ 0.0116 & \textbf{0.3042 $\pm$ 0.0080} & \textbf{0.3372 $\pm$ 0.0096} \\
\textbf{\MyModel-S} & 0.2170 $\pm$ 0.0077 & 0.3237 $\pm$ 0.0069 & 0.3582 $\pm$ 0.0035 & 0.1680 $\pm$ 0.0083 & 0.2906 $\pm$ 0.0054 & \underline{0.3342 $\pm$ 0.0037} \\
\textbf{\MyModel$^\dag$-P} & \underline{0.2232 $\pm$ 0.0039} & \textbf{0.3301 $\pm$ 0.0003} & \underline{0.3948 $\pm$ 0.0006} & 0.1606 $\pm$ 0.0069 & 0.2745 $\pm$ 0.0060 & 0.2835 $\pm$ 0.0028 \\
\textbf{\MyModel$^\dag$-S} & \textbf{0.2321 $\pm$ 0.0015} & \underline{0.3290 $\pm$ 0.0000} & 0.3560 $\pm$ 0.0006 & 0.1650 $\pm$ 0.0024 & 0.2842 $\pm$ 0.0087 & 0.3131 $\pm$ 0.0063 \\
\bottomrule
\end{tabular}
\par\smallskip
{\raggedright\footnotesize $^\dag$ denotes few-shot marker extraction; unmarked \MyModel~rows use zero-shot marker extraction.\par}
\end{table*}

Table~\ref{tab:m-rag-results} reports end-to-end QA performance on three LongBench subtasks under two retrieval backends and three evidence budgets. Across the 18 dataset--retriever--budget settings, at least one \MyModel~variant ranks first or second in 16 settings. The gains are most visible under tight budgets, where the retrieved unit has a direct effect on whether the generator receives answer-bearing evidence. With BAAI/bge-m3 at 128$\times$1, \MyModel~ranks first on Qasper and 2WikiMultihopQA and second on NarrativeQA. With BM25, \MyModel~achieves the best score on NarrativeQA and Qasper at 128$\times$1 and becomes best on 2WikiMultihopQA at 128$\times$3.

The results also show that the benefit is not tied to a single retrieval backend. Dense retrieval benefits from matching queries against compact semantic keys, while BM25 benefits from a concise lexical field that avoids term dilution in long chunks. This consistency supports the main design choice of \MyModel: retrieval should operate over a query-facing representation, while the generation budget should be spent on the paired evidence value.

The advantage is smaller when the evidence budget increases. Strong span-based methods such as PIC and DOS can recover more context at larger budgets and lead in several high-budget settings, including Qasper under dense retrieval and 2WikiMultihopQA under dense retrieval at 128$\times$5. These exceptions are informative rather than contradictory: chunks can work well when enough context can be returned, but \MyModel~is strongest when evidence selection is constrained. We therefore first inspect the constructed semantic index, then analyze its retrieval behavior and robustness.

\subsection{Semantic Index Analysis}
\label{subsec:semantic-index}

\begin{figure}[t]
\centering
\includegraphics[width=\linewidth]{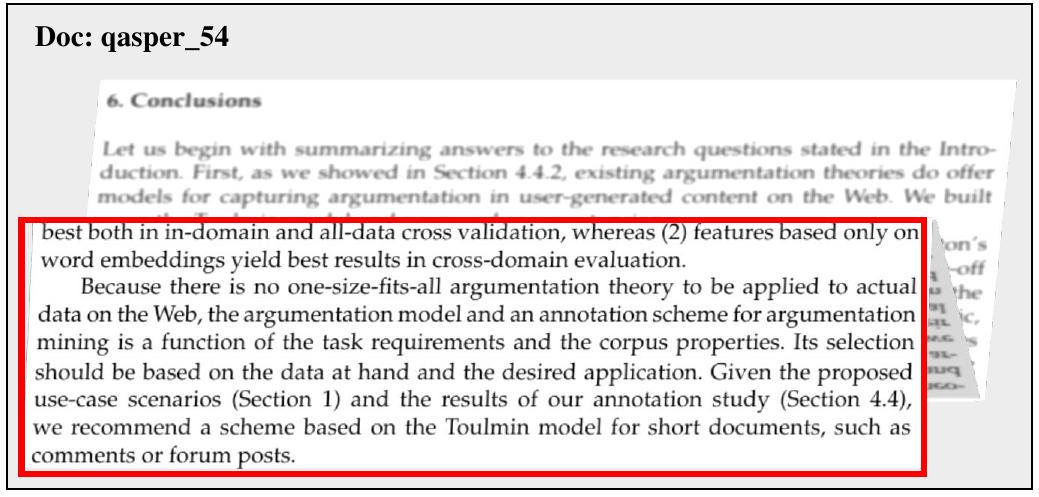}
\caption{A fallback example from document Qasper\_54. Red box highlights the retrieved content.}
\label{fig:fallback_case}
\end{figure}

\noindent \textbf{Document Coverage.} Table~\ref{tab:coverage-markers} reports the provenance coverage of extracted markers. Both zero-shot and few-shot extraction cover more than 99.8\% of source units on average across the three benchmarks, while fallback records account for less than 1\% of the units. This suggests that LLM-based marker extraction does not introduce systematic coverage gaps under the evaluated settings. Figure~\ref{fig:fallback_case} shows a fallback example from Qasper\_54~\cite{habernal2017argumentation}: the uncovered unit comes from a conclusion paragraph with little standalone semantic content. Few-shot extraction slightly reduces fallback usage in NarrativeQA and Qasper, but the zero-shot setting already provides comparable mean coverage, indicating that the schema-level extraction instruction transfers across the evaluated workloads.

\begin{table}[htbp]
\centering
\caption{Statistics on average coverage rates under zero-shot (ZS) and few-shot (FS) prompting.}
\label{tab:coverage-markers}
\resizebox{\linewidth}{!}{
\begin{tabular}{l c c c c}
\toprule
\textbf{Benchmark} & \textbf{Prompting} & \textbf{Avg. Cover.} & \textbf{Variance} & \textbf{Fallback~(\%)} \\
\midrule
\multirow{2}{*}{NarrativeQA} & ZS & 99.80 & 0.4040 & 0.7 \\
 & FS & 99.87 & 0.1948 & 0.2 \\
\midrule
\multirow{2}{*}{Qasper} & ZS & 99.93 & 0.2316 & 0.9 \\
 & FS & 99.90 & 0.2776 & 0.1 \\
\midrule
\multirow{2}{*}{2WikiMultihopQA} & ZS & 99.80 & 0.5379 & 0.0 \\
 & FS & 99.84 & 0.3648 & 0.5 \\
\bottomrule
\end{tabular}
}
\end{table}

\noindent \textbf{Marker Granularity.} Table~\ref{tab:index_distribution} shows that most markers correspond to single-paragraph evidence, indicating alignment with localized discourse units. At the same time, 29.7--36.5\% of markers span continuous multi-paragraph regions, which allows a record to aggregate evidence that crosses a paragraph boundary. Non-consecutive markers are rare, suggesting that \MyModel~mostly preserves the local structure of the source document while still allowing occasional long-range aggregation.

\begin{table}[hbt!]
\centering
\caption{Paragraph index distribution of extracted markers.}
\resizebox{\linewidth}{!}{
\begin{tabular}{l c c c c}
\toprule
\multirow{2}{*}{\textbf{Benchmark}} & \multirow{2}{*}{\textbf{Total}} & \textbf{Single} & \multicolumn{2}{c}{\textbf{Multi-Paragraph}} \\
\cmidrule(l){4-5}
& & \textbf{Paragraph} & \textbf{Continuous} & \textbf{Non-consec.} \\
\midrule
NarrativeQA & 61,632 & 40,755 (66.1\%) & 20,849 (33.8\%) & 28 (0.05\%) \\
2WikiMultihopQA & 16,345 & 11,459 (70.1\%) & 4,856 (29.7\%) & 30 (0.18\%) \\
Qasper & 9,370 & 5,888 (62.8\%) & 3,422 (36.5\%) & \textbf{60 (0.64\%)} \\
\bottomrule
\end{tabular}%
}
\label{tab:index_distribution}
\end{table}

\begin{figure}
\centering
\includegraphics[width=0.95\linewidth]{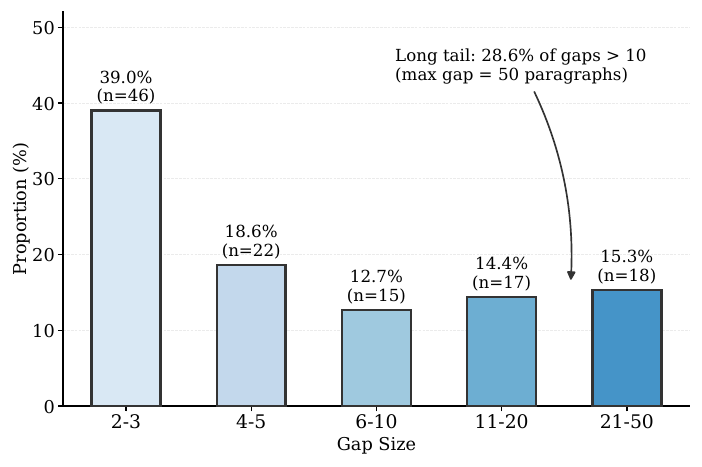}
\caption{Distribution of paragraph index gaps in non-consecutive multi-index markers.}
\label{fig:gap_distribution}
\end{figure}

For the small number of non-consecutive markers, Figure~\ref{fig:gap_distribution} shows the distribution of paragraph gaps. Many gaps remain local, but the distribution has a long tail: 28.6\% of gaps exceed 10 paragraphs, and the maximum span reaches 50 paragraphs. These cases show that marker extraction can occasionally group distant but semantically related evidence, a pattern that fixed local windows are unlikely to capture directly.

\begin{figure}[t]
\centering
\includegraphics[width=\linewidth]{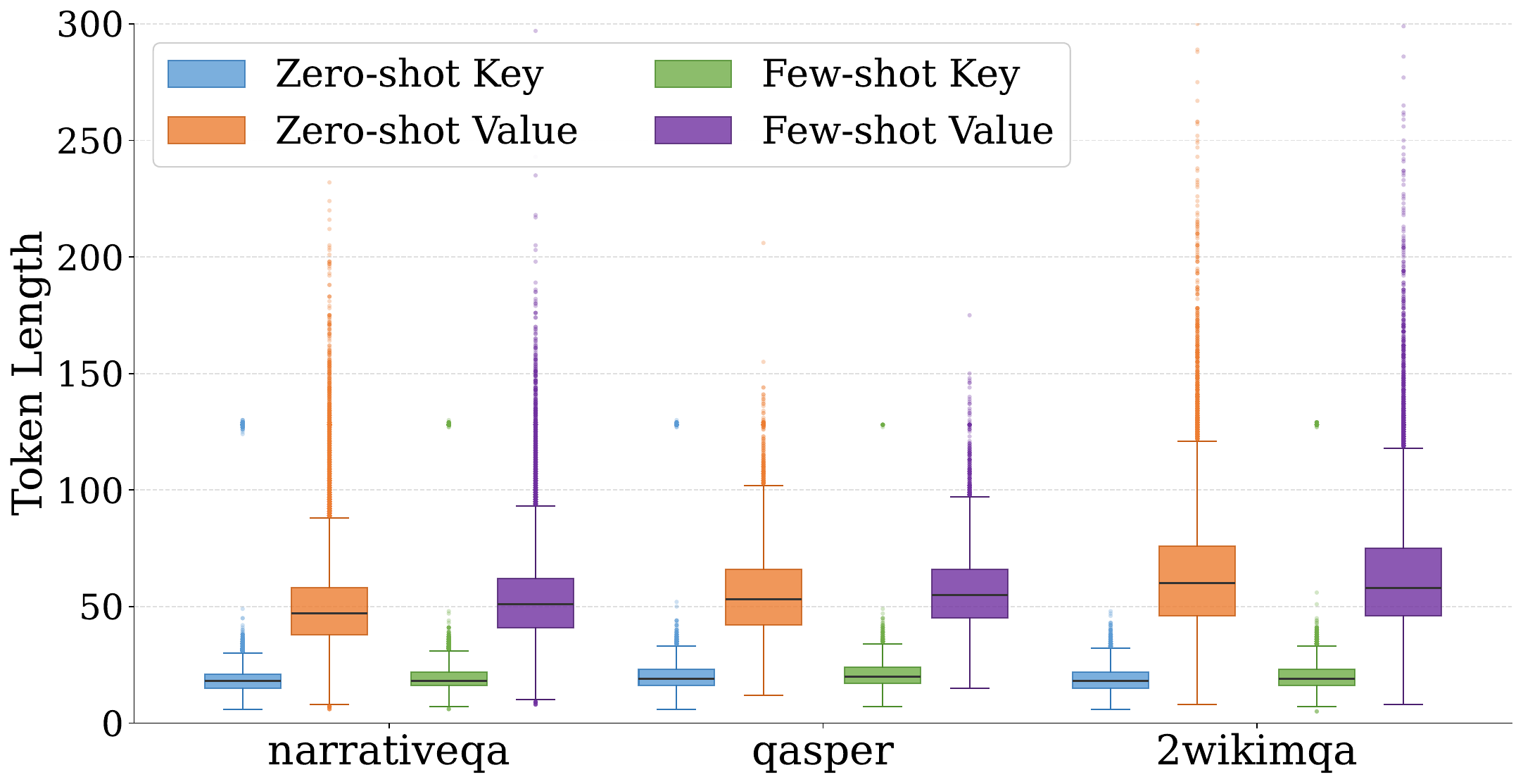}
\caption{Token length of $k, v$ across benchmarks, where boxes show quartiles, whiskers extend to 1.5 IQR, and outliers are marked.}
\label{fig:token-distribution-analysis}
\end{figure}

\noindent \textbf{Token Length of $k, v$.} Figure~\ref{fig:token-distribution-analysis} presents the token length statistics of \textit{retrieval keys} and \textit{information values} under zero-shot (ZS) and few-shot (FS) prompting. Across all benchmarks and prompting regimes, \textit{retrieval keys} remain compact, averaging 19--20 tokens with low variance. \textit{Information values} are longer, averaging 50--65 tokens, or about $2.5$--$3\times$ the length of keys. This length asymmetry is consistent with the intended key-value separation: $k$ serves as a lightweight retrieval representation, while $v$ carries more context for generation. The value length also varies by workload. NarrativeQA tends to produce shorter values, Qasper requires longer explanatory values, and 2WikiMultihopQA yields the longest values, consistent with its multi-hop and cross-document evidence. The compactness of $k$ remains stable across these workloads and across ZS/FS prompting.

\subsection{Retrieval Behavior}
\label{subsec:retrieval-behavior}

This section examines how the semantic key-value index changes the retrieval space and the evidence returned to the generator.

\begin{figure*}[t]
\centering
\includegraphics[width=\linewidth]{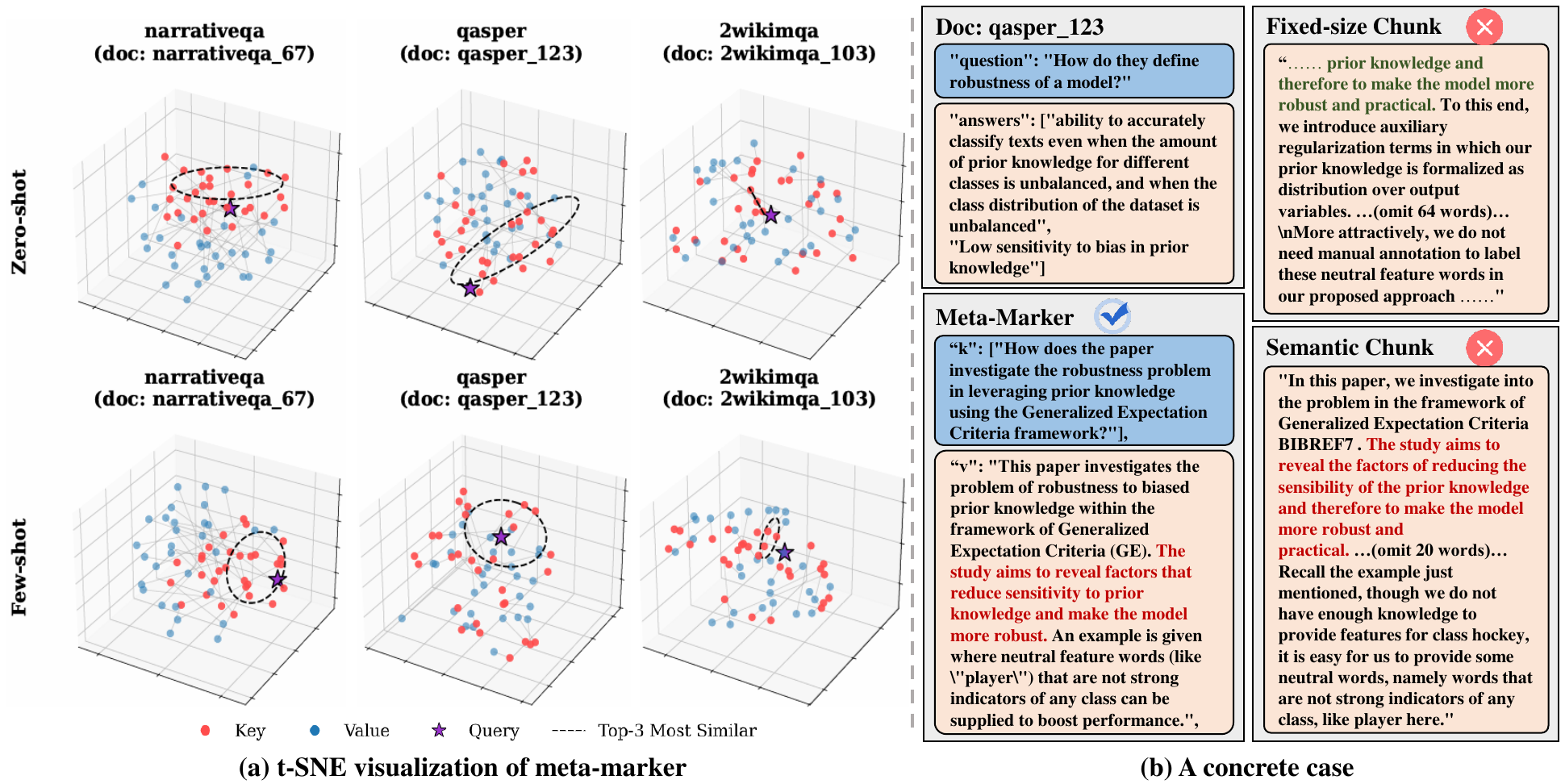}
\caption{Retrieval behavior of \MyModel. The visualizations show query-aligned key neighborhoods and broader value distributions; the Qasper case illustrates how key-based lookup returns a definition-aligned value while chunk-based retrieval mixes multiple themes.}
\label{fig:case}
\end{figure*}

\noindent \textbf{Retrieval Geometry and Case Study.} We conduct case studies using sampled documents from each benchmark. Figure~\ref{fig:case}(a) visualizes the query and corresponding markers using t-SNE. A consistent pattern emerges: \textit{retrieval keys} form localized, query-aligned clusters, whereas \textit{information values} occupy a broader region of the embedding space. This separation holds across benchmarks, including 2WikiMultihopQA where multi-hop reasoning is inherently complex. Few-shot prompting further tightens key clusters, consistent with the reduced variance observed in token length analysis. Figure~\ref{fig:case}(b) presents a Qasper example involving a definition query about ``robustness''. The desired evidence is a sentence-level conceptual statement, not a broader procedural description. Fixed-Size and Semantic retrieval return information-rich passages, but these chunks mix several themes, including regularization, examples, and reference details, and do not directly state the definition. One plausible reason is that chunk embeddings represent blended topics when a span contains multiple semantic units. In contrast, meta-marker retrieval matches the query to an intent-aligned key and returns a value that explicitly encodes the definitional statement, as highlighted in Figure~\ref{fig:case}.

\noindent \textbf{Effect of Key-Value Decomposition.} These observations help explain the behavior of \MyModel~in the main results. At the retrieval stage, compact \textit{retrieval keys} provide a focused lookup field rather than forcing the retriever to search over heterogeneous generation payloads. In dense retrieval, this focus is reflected in the tighter neighborhoods shown in Figure~\ref{fig:case}(a). In sparse retrieval, the same key field exposes concise lexical evidence for BM25 matching, reducing term dilution from long chunks that mix multiple topics. At the generation stage, the length asymmetry between \textit{retrieval keys} and \textit{information values} allows $v$ to retain contextual evidence, while $k$ remains lightweight for efficient lookup. This index-data separation follows a common information-systems principle: the object used for lookup does not need to be identical to the payload returned to the downstream operator. We next test whether this focused lookup field remains selective when the candidate corpus grows.

\subsection{Retrieval Robustness}
\label{sec:corpus-size-robustness}

\begin{figure}[t]
    \centering
    \includegraphics[width=\linewidth]{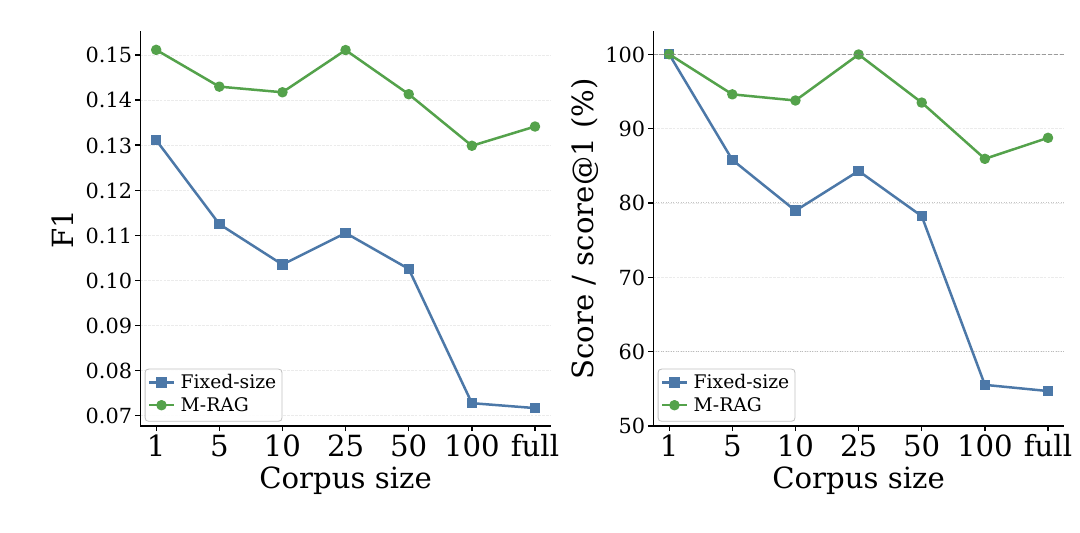}
    \vspace{-6mm}
    \caption{Robustness under expanding candidate corpora on NarrativeQA. \MyModel~retains a larger fraction of its single-document performance as distractor documents are added.}
    \label{fig:corpus-size-robustness}
\end{figure}

To evaluate the robustness of \MyModel~under expanding retrieval spaces, we conduct a controlled corpus-size stress test. For each query, we fix the query, answer generation model, retrieval budget, and the corresponding gold document, while progressively increasing the number of candidate indexed documents available to the retriever. The candidate corpus consists of the gold document together with deterministically sampled distractor documents from the same benchmark. Corpus size is varied from 1 to 5, 10, 25, 50, 100, and finally the full corpus.

Figure~\ref{fig:corpus-size-robustness} reports both the absolute F1 score and the performance retention under different corpus sizes. The retention ratio is defined as the F1 score at a given corpus size normalized by the F1 score when only the gold document is available. As the retrieval space expands from a single document to the full corpus, \MyModel~retains performance better than the fixed-size dense retrieval baseline. On NarrativeQA, the F1 score of \MyModel~decreases from 0.1511 to 0.1341 as the retrieval space expands from a single document to the full corpus, corresponding to a retention ratio of 88.8\%. In contrast, the fixed-size baseline drops from 0.1311 to 0.0717, retaining only 54.7\% of its original performance.

These results indicate that \MyModel~is less sensitive to distractor documents in this stress test. As corpus size grows, the gap between \MyModel~and the fixed-size baseline widens, suggesting that compact semantic keys provide a more selective access path when the retriever must search over a larger candidate space. We then examine whether this behavior depends on prompt wording or context assembly order.

\subsection{Design Sensitivity}
\label{subsec:design-sensitivity}

\noindent \textbf{Prompt Sensitivity.} To check whether marker extraction depends on a particular prompt wording, we evaluate three prompt variants on Qasper under few-shot position sorting. Prompt A is the original extraction prompt, Prompt B swaps the output-format and critical-requirement sections, and Prompt C moves the key-goal section below the critical-requirement section. Table~\ref{tab:prompt-sensitivity} shows that the results remain close across budgets, suggesting that extraction mainly depends on the record schema and coverage constraints rather than a single prompt ordering.

\begin{table}[htbp]
\centering
\caption{Prompt sensitivity on Qasper under few-shot position sorting.}
\label{tab:prompt-sensitivity}
\resizebox{\linewidth}{!}{
\begin{tabular}{lccc}
\toprule
\textbf{Token Budget} & \textbf{Prompt A} & \textbf{Prompt B} & \textbf{Prompt C} \\
\midrule
128$\times$1 & 0.1806 $\pm$ 0.0043 & 0.1866 $\pm$ 0.0004 & 0.1734 $\pm$ 0.0003 \\
128$\times$2 & 0.2210 $\pm$ 0.0018 & 0.2226 $\pm$ 0.0006 & 0.2105 $\pm$ 0.0022 \\
128$\times$3 & 0.2370 $\pm$ 0.0037 & 0.2370 $\pm$ 0.0004 & 0.2318 $\pm$ 0.0000 \\
128$\times$4 & 0.2612 $\pm$ 0.0056 & 0.2534 $\pm$ 0.0009 & 0.2571 $\pm$ 0.0010 \\
128$\times$5 & 0.2693 $\pm$ 0.0061 & 0.2560 $\pm$ 0.0019 & 0.2711 $\pm$ 0.0006 \\
\bottomrule
\end{tabular}
}
\end{table}

\begin{figure*}[htbp]
\centering
\includegraphics[width=0.9\linewidth]{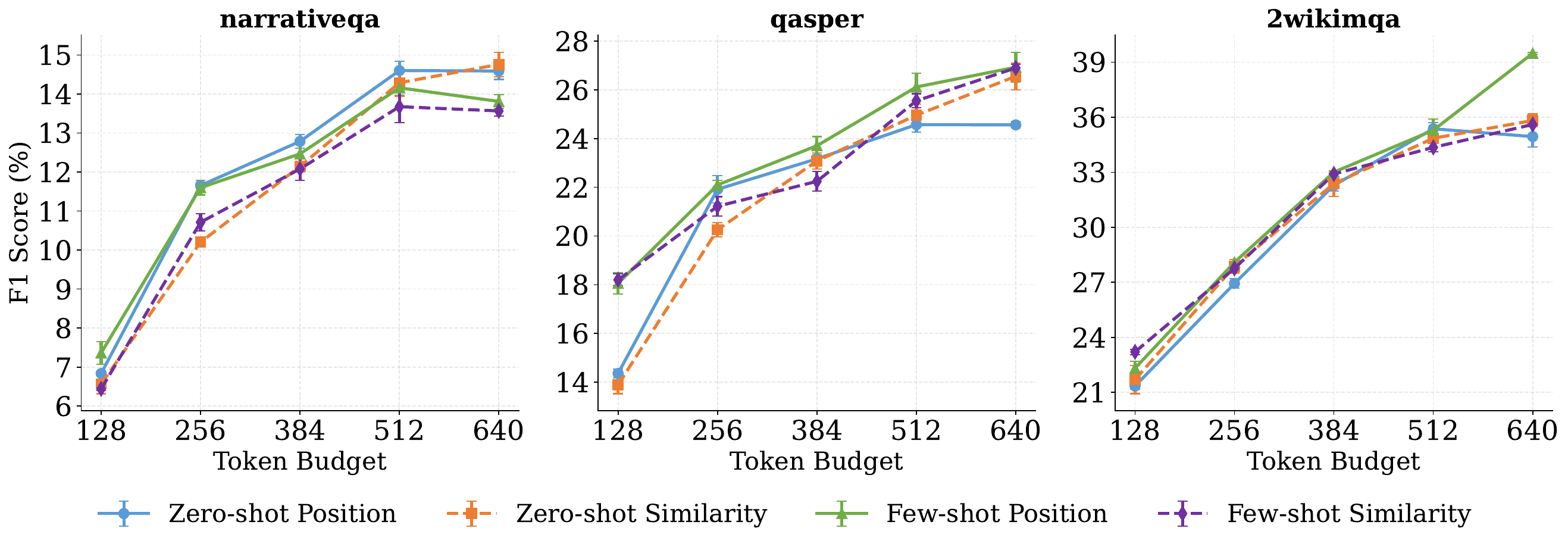}
\caption{Performance comparison of sorting strategies and prompting regimes across benchmarks (token budgets: 128$\times$1, 128$\times$2, 128$\times$3, 128$\times$4, 128$\times$5).}
\label{fig:sorting-analysis}
\end{figure*}

\noindent \textbf{Sorting Strategies.} To examine how context assembly affects \MyModel, we compare position-based sorting (PS) and similarity-based sorting (SS) under zero-shot and few-shot marker extraction. Figure~\ref{fig:sorting-analysis} reports F1 scores across the three benchmarks and five retrieval budgets. Position sorting is generally stronger on structure-sensitive workloads. In 2WikiMultihopQA, PS improves the average score over SS by 0.9\%, with few-shot PS reaching 39.48\% at 640 tokens compared with 35.60\% for SS. A similar pattern appears in Qasper at mid-to-high budgets, where preserving document order helps maintain the structure of technical evidence.

Similarity sorting is more useful for open-ended narrative queries. On NarrativeQA, zero-shot SS reaches 14.76\% at 640 tokens, a 1.2\% gain over PS. This suggests that, when evidence is less tied to document order, prioritizing query relevance during assembly can be more effective. The two sorting policies therefore act as alternative execution choices over the same semantic key-value index: PS favors source structure, while SS favors local query relevance.

\subsection{Efficiency and Cost}
\label{subsec:efficiency-cost}

This section reports online matching time, the efficiency comparison with LightRAG, and the offline costs associated with marker extraction and index maintenance.

\begin{figure}[htbp]
\centering
\includegraphics[width=\linewidth]{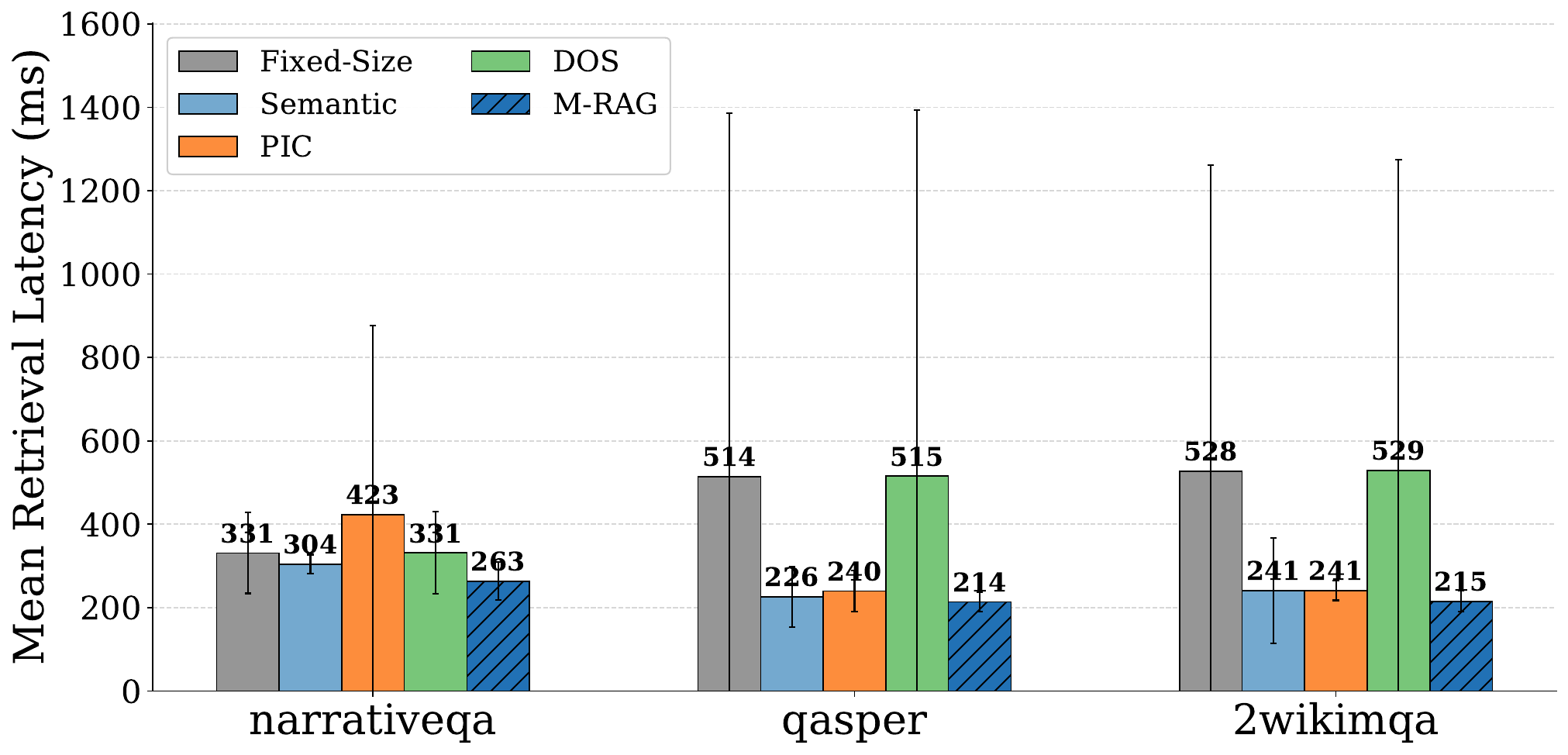}
\vspace{-4mm}
\caption{Online retrieval latency across benchmarks. \MyModel~matches queries against compact semantic keys and achieves lower matching time than chunk-based retrieval units.}
\label{fig:efficiency-analysis}
\end{figure}

\noindent \textbf{Online Retrieval Efficiency.} Figure~\ref{fig:efficiency-analysis} compares online similarity matching time between user queries and retrieval units. \MyModel~has the lowest retrieval latency across all benchmarks. Since retrieval is performed over precomputed dense vectors with the same embedding dimensionality, the difference is not simply a result of shorter text strings at query time. A likely contributing factor is that query-like keys form a more focused retrieval space than raw chunks, whose embeddings may mix several topics.

\begin{figure}[t]
    \centering
    \includegraphics[width=0.9\linewidth]{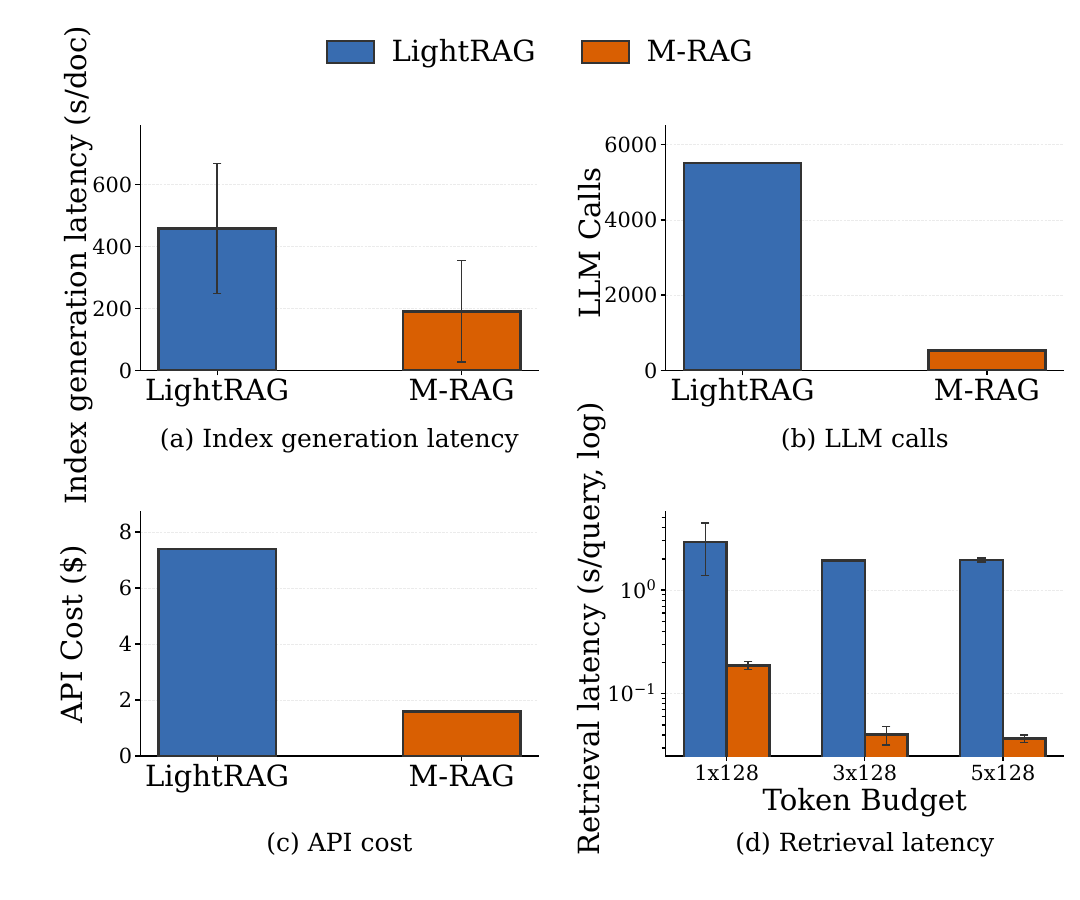}
    \vspace{-4mm}
    \caption{Efficiency comparison between LightRAG and M-RAG. (a) Index generation latency per document. (b) LLM calls required for index construction. (c) Estimated API cost for index construction. (d) Retrieval latency under different retrieval budgets. Error bars denote standard deviation where applicable.
    }
    \label{fig:lightrag_mrag_generation_retrieval_latency}
    \vspace{-2mm}
\end{figure}

\noindent \textbf{Index Construction Efficiency.}
Figure~\ref{fig:lightrag_mrag_generation_retrieval_latency} compares the overall efficiency of LightRAG and \MyModel. Beyond online retrieval latency, we additionally evaluate the cost of offline index construction. Despite generating both retrieval keys and information values, \MyModel~requires approximately 190 seconds per document for index construction, compared with about 460 seconds per document for LightRAG. \MyModel~also has lower retrieval latency across all retrieval budgets in this comparison. Even under a retrieval budget of 5$\times$128 tokens, retrieval latency remains below 0.2 seconds, whereas LightRAG requires several seconds per query. These results show that, relative to this graph-enhanced baseline, semantic key-value indexing offers a lower-cost offline construction path and faster online retrieval.

To further quantify the LLM overhead during index construction, we compare the indexing cost of \MyModel with LightRAG using its default indexing configuration as the baseline. The reported statistics include only the LLM calls required for offline index construction, excluding embedding computation, retrieval, and answer generation. As shown in Fig.~\ref{fig:lightrag_mrag_generation_retrieval_latency}, \MyModel~requires only 525 LLM invocations, compared with 5,507 for LightRAG, representing a 10.5$\times$ reduction in LLM invocations. Although omitted from the figure for clarity, \MyModel~consumes only 7.77M input/output tokens, compared with 42.42M for LightRAG, reducing the estimated API cost from \$7.39 to \$1.59, corresponding to a 4.65$\times$ reduction in indexing cost. This improvement stems from \MyModel's document-level semantic marker extraction, which generates semantic markers once per document, whereas LightRAG repeatedly performs chunk-level entity extraction across the corpus.

\noindent \textbf{Amortized Offline Cost.}
Practical deployment of \MyModel~requires considering the cost of LLM-based marker extraction, but this cost belongs to the offline index-construction phase rather than online query processing. Using the DeepSeek-V3.2 API with \textit{Context Caching}, processing 16.1M tokens costs about \$0.56 per 1M tokens ($\approx$ \$0.053 per 1,000 markers). Although sequential extraction takes about 17.3 minutes per document, marker extraction is naturally parallel across documents; with moderate concurrency, e.g., 100 requests, generating 1M output tokens requires 0.209 hours.

Because marker extraction is performed once during offline index construction, its cost is amortized over all subsequent queries served by the same index. Let $C_{\mathrm{ext}}$ denote the one-time LLM-based marker extraction cost, $C_{\mathrm{idx}}$ the one-time key embedding and indexing cost, and $C_{\mathrm{ret}}$ the online retrieval and value lookup cost per query. For $Q$ queries over a fixed index, the effective non-generation cost per query is
\begin{equation}
    C_{\mathrm{eff}}(Q) = \frac{C_{\mathrm{ext}} + C_{\mathrm{idx}}}{Q} + C_{\mathrm{ret}} .
\end{equation}
As $Q$ increases, the amortized offline term decreases monotonically and the effective cost approaches the online retrieval cost, i.e., $\lim_{Q\to\infty} C_{\mathrm{eff}}(Q)=C_{\mathrm{ret}}$. This formulation captures the write-once, read-many behavior of \MyModel~without requiring a separate amortization experiment.

\noindent \textbf{Incremental Updates for Dynamic Knowledge Bases.}
For dynamic knowledge bases, \MyModel~does not need to rebuild the entire marker index when new documents arrive. Instead, only the newly added documents are processed to extract markers, embed keys, and append entries to the existing index. If the existing collection contains $N$ documents and an update adds $\Delta N$ documents, full rebuilding scales with $N+\Delta N$, whereas incremental maintenance scales only with $\Delta N$:
\begin{equation}
    C_{\mathrm{update}}^{\mathrm{inc}} = O(\Delta N), \qquad
    C_{\mathrm{update}}^{\mathrm{full}} = O(N+\Delta N).
\end{equation}
This follows directly from the independence of offline marker extraction across documents: previously extracted markers and paired values remain unchanged, while new entries can be appended to the key-value index.

\subsection{Limitations}
\label{subsec:limitations}

\begin{table}[hbt!]
\centering
\caption{A failure case from 2WikiMultihopQA.}
\label{tab:failure_cases}
\small
\begin{tabular}{@{}p{\columnwidth}@{}}
\toprule
\textbf{Case Study: Granularity Mismatch} \\
\midrule
\textbf{Query:} When is Henrietta Maria Of Brandenburg-Schwedt's father's birthday? \\
\textbf{\textit{Retrieval Key} ($k$):} What were Philip William's titles and his role as governor? \\
\textbf{\textit{Information Value} ($v$):} \textit{Philip William, Prince in Prussia (19 May 1669--19 December 1711) was the first owner of the Prussian secundogeniture...} \\
\bottomrule
\end{tabular}
\vspace{-2mm}
\end{table}

Table~\ref{tab:failure_cases} illustrates a failure case of \MyModel. The $k$-$v$ decomposition preserves the answer in the \textit{information value}, but retrieval can still fail when the \textit{retrieval key} is too coarse for the query. In this example, the key summarizes Philip William's titles and governance roles, whereas the query asks for a specific relational attribute: the father's birthday. Although the correct date appears in the value, the key does not expose that relation clearly enough, weakening the $q$-$k$ match during retrieval. This limitation is most likely for fine-grained relational questions, where a generated key may summarize an entity or event without naming the exact relation needed by the query. The failure therefore comes from query--key misalignment rather than loss of source evidence.

\section{Conclusion}
\label{sec:conclusion}
This paper studies RAG as a data access problem under a token budget. We show that chunking couples two different roles: the record used for lookup and the evidence payload consumed by the generator. \MyModel~addresses this coupling by materializing documents into semantic key-value records, where compact retrieval keys support dense or sparse lookup, information values provide faithful generation evidence, and provenance pointers make the index traceable and order-aware. The index is constructed offline with coverage validation and fallback repair, and queried online by retrieving keys and assembling paired values within the context budget. Experiments on LongBench QA subtasks show that \MyModel~is especially effective when evidence budgets are tight, while further analyses show high source coverage, robustness to larger candidate corpora, and lower retrieval latency. These results support a simple design principle: the object used for lookup need not be the object returned for generation. Semantic key-value indexing therefore provides a modular access layer for RAG systems, leaving room for stronger key refinement, provenance-aware execution plans, and incremental index maintenance.

\section*{AI-Generated Content Acknowledgement}
In the preparation of this manuscript, we used Claude Code to assist with programming, Nano Banana~\cite{comanici2025gemini25pushingfrontier} to generate Figure~\ref{fig:chunking}, and ChatGPT for grammar polishing. The authors reviewed and verified all AI-assisted content and remain responsible for the correctness and originality of the manuscript.

\bibliographystyle{IEEEtran}
\bibliography{Reference}



\end{document}